\documentclass[lettersize,journal]{IEEEtran}
\usepackage{amsmath,amsfonts}
\usepackage{array}
\usepackage{textcomp}
\usepackage{stfloats}
\usepackage{url}
\usepackage{verbatim}
\usepackage{graphicx}
\usepackage{cite}

\usepackage{textcomp}
\usepackage{xcolor}
\usepackage{algpseudocode}
\usepackage{caption}
\usepackage{subcaption}
\usepackage{subfloat}
\usepackage{float}
\usepackage[linesnumbered,ruled]{algorithm2e}
\usepackage{enumitem}
\usepackage{fancybox}
\usepackage{multirow}
\usepackage{booktabs}

\begin{document}

\title{TPPD: Targeted Pseudo Partitioning based Defence for Cross-Core Covert Channel Attacks}

\author{Jaspinder Kaur, and Shirshendu Das~\IEEEmembership{~IEEE Member}
\thanks{The authors are associated with the Department of CSE, Indian Institute of Technology Ropar, Punjab, India, 140001. Email: \{2017csz0002@iitrpr.ac.in, shirshendu@iitrpr.ac.in\}}
}



\maketitle

\begin{abstract}
Contemporary computing employs cache hierarchy to fill the speed gap between processors and main memories. In order to optimise system performance, Last Level Caches(LLC) are shared among all the cores. Cache sharing has made them an attractive surface for cross-core timing channel attacks. In these attacks, an attacker running on another core can exploit the access timing of the victim process to infiltrate the secret information. One such attack is called cross-core Covert Channel Attack (CCA). Timely detection and then prevention of cross-core CCA is critical for maintaining the integrity and security of users, especially in a shared computing environment. In this work, we have proposed an efficient cross-core CCA mitigation technique. We propose a way-wise cache partitioning on targeted sets, only for the processes suspected to be attackers. In this way, the performance impact on the entire LLC is minimised, and benign applications can utilise the LLC to its full capacity. We have used a cycle-accurate simulator (gem5) to analyse the performance of the proposed method and its security effectiveness. It has been successful in abolishing the cross-core covert timing channel attack with no significant performance impact on benign applications. It causes $23\%$ less cache misses in comparison to existing partitioning based solutions while requiring $\approx0.26\%$ storage overhead.
\end{abstract}

\begin{IEEEkeywords}
Cache Security, Timing Channel Attacks, Cache Partitioning, Covert Channel Attack (CCA), Last Level Cache (LLC).
\end{IEEEkeywords}

\section{Introduction}
\label{sec:intro}
\begin{figure*}[]
	\centering
	\includegraphics[scale=0.5]{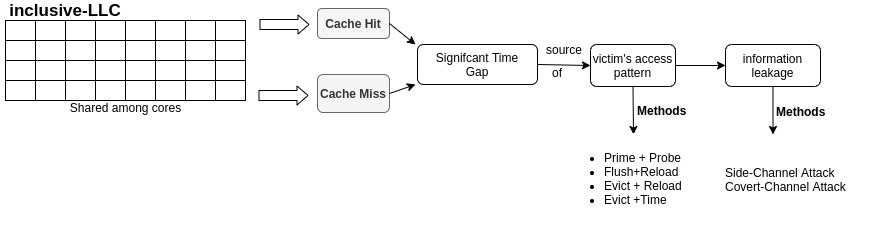}
	\caption{The step-wise details of LLC based timing channel attacks}
	\label{fig:blockDiagramCacheTimingChannelAttack}
\end{figure*}

Modern multi-core processors have adopted a multi-level cache hierarchy to address the rising needs for high-performance computing. Caches are small-sized faster memories deployed to bridge the speed gap between processor and main memory. Most commercial processors available in the market are equipped with multiple levels of caches. The top levels of caches are private to each core, and the Last Level Cache (LLC) is shared among all the processing cores\footnote{All the cache memories in this paper are considered as set-associative cache. We used the term ``set'' and ``way'' of a set-associative cache without detailed explanation about them.}. Cache memories boost overall performance by allowing processors quicker access to data. While improving the system's overall performance, the LLCs have also become an attractive target surface for timing channel-based attacks \cite{percival} due to these reasons:

\begin{enumerate}
	\item The significant time difference between a cache hit and miss provides an effective timing channel that is exploited to unveil the memory accesses of the targeted process \cite{efficient,cca-attack}.
	\item Shared nature of LLC allows an attacker process to interfere in cache occupancy of other processes. The attacker uses the timing channel to understand other processes' access patterns on the shared cache as described in Figure \ref{fig:blockDiagramCacheTimingChannelAttack}. These accesses are further analysed to reveal the underlying secret \cite{cache-time}.
\end{enumerate}

Any remote process, sharing LLC with the victim process, can mount these attacks without requiring special privileges or shared address space \cite{lastlevelcachesidechannels}. The victim and attacker process may run in separate cores, but they share the LLC space. Such attacks are called \textit{cross-core attack}. These attacks become more threatening in shared computing environments such as the cloud, where multiple users from different security domains share the underlying LLC. Cache timing channel can be implemented in two forms: Side-Channel Attack(SCA) and Covert Channel Attack(CCA) \cite{survey-jas,survey-prabhat}. The private keys of cryptography algorithms like AES \cite{aes}, RSA \cite{rsa}, and ECDSA \cite{ecdsa} have been the primary target of cache-based SCAs. In a CCA, two suspicious processes, spy and trojan, communicate covertly by exploiting the cache timing channel \cite{survey-jas}. The trojan runs on a different core and knows some secret information that it needs to send to the spy. The attacker uses CCA when no direct communication between trojan and spy is possible because of the security restrictions. 
It is crucial to detect and prevent these attacks effectively in order to preserve the confidentiality and integrity of the computer systems \cite{tracedriven,cachegames}. This paper is based on proposing countermeasures for preventing cross-core CCA attacks. Section \ref{sec:back-cca} discusses CCA in more detail. 

The CCA attack that is discussed in this paper is cross-core attacks; based on the shared LLC used by the multi-core processors. Attack defence mechanisms based on not allowing cache sharing through LLC partitioning have been proposed previously \cite{FairSDP,cotsknight}. Cache partitioning techniques were initially proposed for improving the performance of the system by fairly dividing the LLC among the cores (or applications) \cite{ucp,partition-halwe,partition-new}. Most of them were not proposed keeping security in mind. These partitioning techniques divide the LLC either way-wise or set-wise. The way-wise partitioning techniques are more prevalent where the ways of a set-associative LLC is partitioned among the cores (or applications) \cite{ucp}. The partition can be either static or dynamic. The static partitioning techniques cannot change the partition during the execution, while the dynamic partitioning techniques can change the partition based on the requirement of the running applications. Since cache partitioning can separate the LLC space used by trojan and spy, the cache interference of two applications can be prevented. However, there are two major challenges in using cache partitioning for preventing CCA:
\begin{enumerate}
	\item The static partition-based solutions suffer from severe performance degradation as cache cannot be utilised fairly based on the dynamic behaviour of the system \cite{ucp}.
	\item Dynamic partitioning techniques utilise the cache more efficiently and hence improve performance. However, the dynamic partitioning technique itself can be exploited to mount CCA as proposed in \cite{anuraag}. In this work, the authors have shown that if the attacker knows the logic of partition change, it can misuse this and change the partition as per the attacker wish. Thus, a dynamic partitioning based countermeasure can restrict the attack but indirectly opens another option for CCA.
\end{enumerate}
Hence for preventing CCA, a partitioning technique is required, in which the attacker cannot change partition size. Also, the technique should not degrade the performance of the overall system.

There are some existing countermeasures proposed based on the static partition where the technique applies to all the sets within the LLC. In this paper, we call this type of countermeasures as non-targeted countermeasures. Such countermeasures cause higher performance degradation in the system. Our attack mitigation mechanism overcomes the drawbacks of the existing partitioning-based solutions. We do this by providing Targeted Pseudo Partitioning based Defence (TPPD). It is termed as ``targeted'' because the attack prevention mechanism is only applied to the cache sets suspected to be participating in CCA. The sets which participate in the timing channel attacks are termed as \textit{targeted set}. The remaining cache sets can behave as before, thus minimising the effect of static partitioning on the performance. A practical attack detection technique is deployed that send the information regarding suspicious processes (trojan and spy) and cache sets involved. Attack detector based on conflict misses pattern \cite{prodact} has been tested to identify the suspicious processes and sets involved. 

The main contributions of this work are listed as following:
\begin{enumerate}
	\item We propose an effective attack defense mechanism, TPPD, that dismantles the cross-core CCA with an insignificant effect on system performance.
	\item The proposed TPPD creates the pseudo partition only on the targeted set and is applied only for the trojan and spy process.
	\item Effect of TPPD on the performance of PARSEC benchmark\cite{parsec} is tested experimentally in gem5 simulator\cite{gem5}.
	\item Experiments are conducted to test the effectiveness of TPPD in mitigating cross-core CCA.
\end{enumerate}

The organization of the paper is as follows. Section \ref{sec:background} describes the background and related work. Section \ref{sec:threat} describes the threat model taken under consideration, and Section \ref{sec:proposed} discusses our proposed work in details. The experimental analysis are shown in Section \ref{sec:exp}. Finally Section \ref{sec:con} concludes the paper.

\begin{figure*}[h]
	\centering
	\includegraphics[scale=0.7]{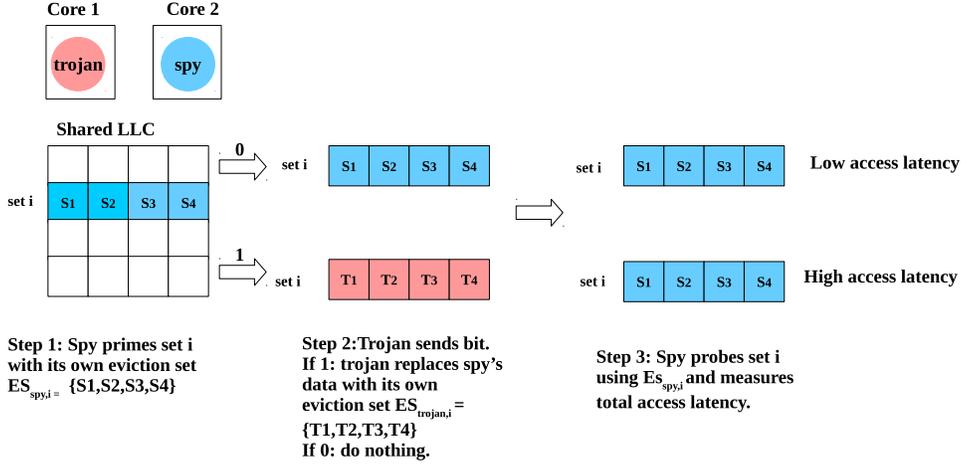}
	\caption{Cross-Core covert channel attack using Prime+Probe technique}
	\label{fig:ppexample}
\end{figure*}

\section{Background and Related Work}
\label{sec:background}
This section describes the background information required to understand the proposed attack defence mechanism. In cache timing channel attack, the attacker process uses different attack methods to unveil the cache access pattern of the target process. In side-channel attacks, the target is an innocent victim process performing cache accesses as part of its underlying operation without any malign intentions. In a covert channel attack, there are two malign processes performing cache accesses with the intent of leaking secrets. An abstract overview of cache timing channel attacks is shown in Figure \ref{fig:blockDiagramCacheTimingChannelAttack}.

\subsection{Cross-Core Covert Channel Attacks}
\label{sec:back-cca}
In a covert timing channel attack, two suspicious processes: spy and trojan, are involved. Trojan has access to some critical information that it wants to transmit to spy, but this transmission is not allowed under system security policy. 
Trojan uses cache timing channel to leak this information to spy without being noticed. This type of attack is called Covert Channel Attack (CCA). The CCA is called cross-core when the spy and trojan execute on two different cores. In this case, both trojan and spy use the shared LLC to create a timing channel. Cache timing channel attack can be constructed using various attack techniques like Prime+Probe (P+P), Evict+Reload (E+R), Flush+Reload (F+R), and Evict+Time (E+T) \cite{survey-jas}. These attack techniques, even though different, follow these basic three steps:

\begin{itemize}
	\item Step 1: In the first step, the aim is to bring the shared LLC in a predictable state known to the attackers. It is done by either bringing some data in LLC or evicting it out by the spy. 
	
	\item Step 2: In this step, the spy process sits idle, and the trojan process performs conditional memory accesses based on the bit to transmit (bit $1$ or $0$) . 
	
	\item Step 3: In this step, the spy process observes the changes made in the earlier known state of the cache by the trojan to detect the bit as $1$ or $0$.
\end{itemize}
Repeating the above mentioned steps, the trojan can send multiple bits to the spy. 

CCA using Prime+Probe (P+P) technique is described in Figure \ref{fig:ppexample}. In this method, spy and trojan do not require any shared address space. Both these processes have their own Eviction Set (ES)\cite{basicevictionset}. ES is the group of unique block addresses mapping to the same set. The number of addresses in each ES is either equal to or more than the associativity of the underlying cache. In the prime phase, the spy uses the addresses present in its ES to fill the targeted set with its own blocks. After priming the cache, spy waits for a trojan to send the bit. If trojan wants to transmit bit $1$, it replaces all of spy's block from the targeted set with its own blocks (from trojan's ES). For bit $0$, it does not do anything; thus, spy's blocks stay in the set. After this, in the probe phase, spy accesses the addresses from its ES again. In this phase, spy faces high latency in case trojan has removed its blocks from the set; otherwise, less latency. Based on this latency, the spy unveils the cache access pattern of trojan and the secret bit transmitted. The rest of the discussions of this paper assume Prime+Probe (P+P) based CCA.

\subsection{Existing Attack Mitigation Techniques}
\label{sec:back-existing}
As discussed in Section \ref{sec:back-cca}, cache sharing is the critical pre-requirement of cross core cache timing channel attacks. Prohibiting cache sharing across processes or different security domains is an effective solution. However, it is not feasible as the impact of a non-shared cache will lead to under-utilisation of cache capacity thus, impacting system performance significantly. Various works with different cache partitioning techniques that try to maintain cache utilisation have been proposed in recent years. Partitioning Locked cache (PLcache)\cite{newcachedesigns} allows compiler and programmer to mark security critical lines, and these will be locked in the cache. In this architecture, each cache line is augmented with process id and a bit dictating whether that line is locked or not. A locked line of a process cannot evict a locked line of a different process, while no unlocked line is allowed to replace a locked line. Thus, disabling inter-process and intra-process cache interference that has been identified as the root cause of cache access-based attacks. NonMonopolizable (NoMo)\cite{nomo} cache allocated $v$ number of ways of each set to an active thread. The allowed range for $v$ is 1 to $A/M$, where $A$ is the associativity of the cache, and $M$ is the maximum number of threads allowed per processing core. Data of oner thread cannot evict reserved lines of another thread; thus, an attacker can not know the victim's cache access. 

NoMo and PLCache offer easy to implement static partition solutions, but the static partitioning can lead to not utilising cache to its full capacity. Security Dynamic Cache Partitioning(SecDCP) cache \cite{secdcp} addresses this issue by proposing a dynamic cache partition. SecDCP allocates a number of ways to different security classes based on their requirement. However, dynamic partitioning in itself can be exploited to create cache covert timing channel attack as described in \cite{anuraag}. A secure dynamic partitioning in terms of SCA is proposed in \cite{FairSDP}. It proposes a secure dynamic cache partitioning cache called FairSDP, where partitioning size is determined by the cache usage of non-critical processes, thus not revealing the requirement of security-critical processes. However, this does not work against CCA as it involves two suspicious processes, and there is no security-critical process. 

All the mitigation techniques discussed above apply some strict cache partitioning policy across all sets for all processes. Hence these techniques reduce the performance of the LLC as well as the entire system. Also, most of the existing countermeasures are non-targeted, as the mechanism is applied to all the cache sets and processes regardless of their participation in the attack. A targeted countermeasure can be less expensive as the necessary actions can be applied only in the sets currently under attack. An alternative solution to prevent cross-core CCA is the recently proposed randomised LLCs \cite{ceaser,ceaser-s}. In these works, the block-to-set mapping of set-associative cache changes after an epoch. However, these techniques suffer significant performance degradation because of the remapping \cite{damru}. The number of write-backs required to remap is high. Also, the policies may not be useful for recent replacement policies like Hawkeye \cite{hawkeye}. Hence in this work, we have been motivated to propose a targeted countermeasure for the cross-core CCA. 

\section{Threat Model}
\label{sec:threat}
In this paper, we have assumed a chip multiprocessor (CMP) with two levels of cache memories. Each core in the CMP has its own private L1 cache, and all the cores share a common L2 cache as LLC. The LLC is inclusive, and all the cache memories are set-associative. The cores, L1 caches, and LLC are connected with some on-chip interconnects. The cross-core CCA attack discussed in this paper runs spy and trojan on two different cores. These two attacking processes (spy and trojan) are not using any shared address space. Hence no cache blocks can be shared among these two processes, and the eviction sets (c.f. Section \ref{sec:back-cca}) of both the processes are also different. However, both spy and trojan share the LLC, and they can evict each other blocks from the LLC. The eviction set of both spy and trojan maps to the same targeted set. We propose TPPD to mitigate cross-core CCA. This attack can be based on Prime+Probe \cite{survey-jas}, Evict+Reload \cite{survey-jas}, or Evict+Time \cite{survey-jas} methods, where spy and trojan processes rely on replacing each other's block to transmit bits. Trojan has a piece of secret information that spy cannot access directly due to underlying system security policies. Trojan transmits this information to spy through LLC based covert communication channel (details of CCA is already discussed in Section \ref{sec:back-cca}). Both trojan and spy can successfully create their eviction sets as they are aware of the virtual address to LLC set mapping. Both attacker processes are unprivileged processes that do not have any kind of special privilege.

\begin{figure*}[ht]
	\centering
	\includegraphics[scale=0.55]{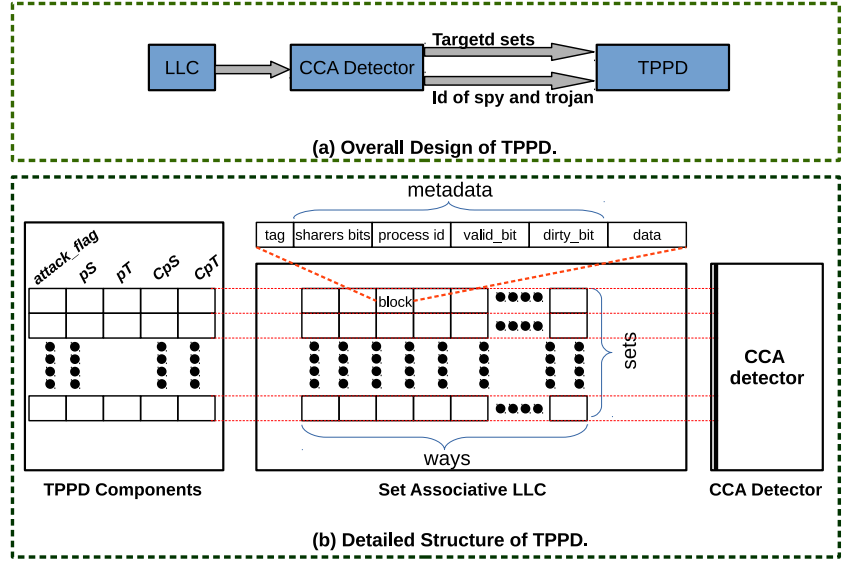}
	\caption{Overall design of TPPD and its structure}
	\label{fig:overallViewTPPD}
\end{figure*}

\begin{figure}[h]
	\centering
	\includegraphics[scale=0.7]{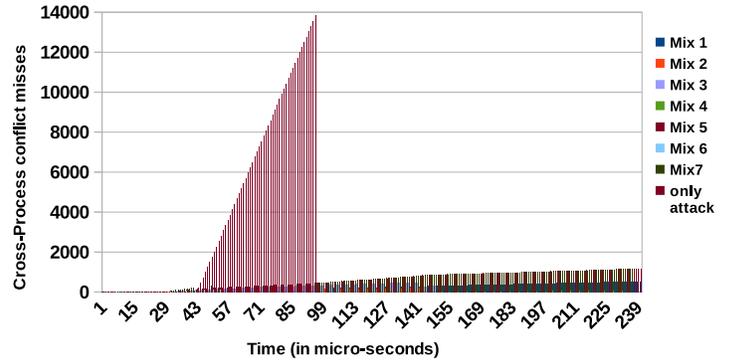} 
	\caption{Cross-process conflict misses per unit of time(.10 seconds).}
	\label{fig:motivation}
\end{figure}

\section{Our Proposal}
\label{sec:proposed}
Our proposed targeted defense mechanism, TPPD, works against LLC based covert channel attacks (cross-core CCA) by creating way-wise pseudo partitioning between the suspicious process pair (spy and trojan). This partition technique used in TPPD cannot be changed by the attacker process as per its requirement. Also, the partition is only applied on the sets suspected to participate in the covert channel communication. The purpose of using the term ``pseudo'' is discussed later in this section. TPPD decreases the cache interference between the trojan and spy on the targeted sets, thus disturbing the access latency pattern observed by the spy to interpret the bit sent. TPPD receives information regarding suspicious process ids (spy and trojan) and the targeted sets from a CCA detector module; then applies the proposed pseudo-partitioning on these sets for the suspicious processes. The CCA detection module is discussed in Section \ref{sec:detection}. The overall design of TPPD is shown in Figure \ref{fig:overallViewTPPD}(a). In this section, we describe TPPD in detail, along with how it can be implemented with minimal modification in architectural design.

\subsection{Cross-Core Covert Channel Attack Detector}
\label{sec:detection}
TPPD works efficiently by applying attack mitigation mechanisms only for suspicious process pairs on the LLC sets suspected to be used for cache covert channel attacks. The countermeasure proposed in TPPD depends on an efficient CCA detection mechanism. In this section, we describe how existing attack detection techniques can be modified or extended to identify suspicious processes and cache sets.

\subsubsection{Conflict Misses Pattern based Detector}
A cross-core CCA detection mechanism based on conflict misses pattern is proposed in \cite{prodact}. It is based on the observation that set under attack have a higher number of conflict misses than innocent processes, and these misses form a ping pong pattern because spy and trojan processes evict each other's data regularly to communicate covertly using cache timing channel attacks. This work proposed a two-step detection mechanism in which, firstly, sets with low conflict misses are filtered out. In the second step, sets observing ping pong pattern of conflict misses between two processes are declared suspicious along with processes participating in this pattern. However, we observed that observing total cross-process misses per unit of time is sufficient in order to identify sets and processes used for mounting CCA. We have also observed the same experimentally as shown in Figure \ref{fig:motivation}. The figure shows the cross-process conflict misses on an LLC set. The experiment is performed on different mixes of benign processes (Mix-$x$) from Parsec benchmarks (described in Table \ref{tab:benchmarkMix}) and also on an attacker process pair (only-attack). The details about the experiment setup, benchmarks and creating attack process pair are discussed in Section \ref{sec:exp}. From the figure, it can be observed that the attacker process pair suffers high cross-process conflict misses when compared to any mix of benign benchmarks for the duration of the attack.

\subsubsection{Cache Occupancy based Detector}
Another way to detect the cache sets and the processes involved in covert channel communication is through tracking cache occupancy, process wise \cite{negative-corelation}. This detection method is based on the observation that cache occupancy of the suspicious process have a high negative correlation as these two keep evicting each other's block frequently. This method requires collecting cache occupancy traces and then further analysing these traces to determine occupancy correlation between processes. It was observed that benign processes do not have such a high negative correlation. This method did not require any additional storage overhead and was able to detect suspicious processes with high accuracy. However, this method was able to disclose the identity of suspicious processes with high accuracy but not the cache sets on which the attack was mounted as cache occupancy was measured for the whole LLC. If cache occupancy trace is collected for each set individually, then suspicious sets can be observed as well.
\begin{center}
\begin{table}[]
\caption{PARSEC benchmark mixes used.}
\label{tab:benchmarkMix}
\centering
\begin{tabular}{  m{2cm} | m{4cm} } \toprule
 Mixes & Benchmarks Involved \\ \midrule
 Mix 1 & blacksholes + canneal \\
 Mix 2 & blacksholes + dedup \\  
 Mix 3 & blacksholes + fluidanimate \\ 
 Mix 4 & blacksholes + freqmine \\ 
 Mix 5 & canneal + dedup \\ 
 Mix 6 & canneal + fluidanimate \\ 
 Mix 7 & freqmine + fluidanimate \\ \bottomrule
\end{tabular}
\end{table}
\end{center}
\subsubsection{The detection mechanism used for TPPD}
In this work, we have used the concept of conflict-miss pattern-based CCA detector to detect the attack. However, the cache occupancy-based detector can also be used to detect the attack. Since our work is on detection based countermeasure, we have used the existing detection techniques. Figure \ref{fig:overallViewTPPD}(a) shows that the outcome of the CCA detector is a pair of attacking processes (trojan and spy) and the targeted sets. 
\subsection{Targeted Pseudo Partitioning based Defense (TPPD)}
\label{sec:TPPD}
When the CCA detector detects an attack, it sends information about the processes and sets involved in the CCA. TPPD enables pseudo partitioning on the suspected sets for the suspected process pair (trojan and spy). Two non-overlapping way-wise partitions are created and assigned to trojan and spy. The spy and trojan are unable to replace each other's data if the block is present in this allocated partition. This restriction, however, exists solely for identified malicious pair; other benign processes can access this cache set as before without any constraint. Because of this reason, the proposed partitioning is called pseudo as it is invisible to all processes except attackers. As our defense mechanism is targeted, it only affects the cache sets and the processes involved in CCA, thus not impacting the system's performance significantly.

The proposed TPPD can handle multiple CCA attacks on different cache sets simultaneously. However, in this section, we have assumed only one pair of attacker processes (spy and trojan) and only one targeted set. For the rest of this section, we assume that the detector detects $pS$ as a spy and $pT$ as a trojan. Here $pS$ and $pT$ are the process id of spy and trojan, respectively. Also, we assume that the detected targeted set is $s'$. TPPD can be implemented on top of any replacement policy with minor modifications. A replacement policy has three major modules: (a) insertion, (b) promotion and (c) eviction \cite{partition-halwe}. To implement TPPD, minimal changes are required in the eviction module of a replacement policy. TPPD expects two victim selection approaches on the targeted set $s'$. We call it as ``dual victim policy''. 
\begin{itemize}
	\item[a)] The default victim selection of a replacement policy. Let us represent the victim block selected through this method as $V(s')$, where $s'$ is the targeted set.
	\item[b)] Victim block selection excluding a particular process $p$. Let us represent the victim block selected through this method as $V_x(s',p)$. Here the victim block is selected from the targeted set $s'$, excluding the blocks owned by process $p$.
\end{itemize}
This dual victim policy can be implemented on most of the existing replacement policies \cite{hawkeye,rrip} with minor modifications. The concept of TPPD, generic to any replacement policy (those can be modified for dual victim selection), is discussed next.

Consider that for a cache block $b$, $O(b)$ represents the owner process to which the block belongs. The dual eviction policy is only required for the targeted set $s'$, detected by the CCA detector as discussed in Section \ref{sec:detection}. For other sets, $s; s \ne s'$, the victim block is always $V(s)$. Since the attacker can target any of the cache sets, the facility of dual victim selection must be present in all the sets. However, it is used only for the targeted set. To implement the dual eviction policy, we need to maintain two counters for each set $CpS(i)$ and $CpT(i)$ where $i$ is the set number. $CpS(s')$ and $CpT(s')$ are used to count the number of spy and trojan blocks respectively present in $s'$. When $pT$ (trojan)  block needs to replace an existing block of $pS$ (spy) from $s'$, TPPD sets some additional conditions. If $O(V(s'))$ is $pS$ (spy) and $CpS(s')$ is less than a threshold, $pT$ cannot replace $V(s')$. In that case, the victim block is selected by $V_x(s',pS)$. This policy restricts the trojan to remove all of the spy blocks from $s'$. A similar policy is also applied when a spy tries to evict a block of trojan from $s'$. The ping-ponging pattern of conflict misses between the spy and trojan is interrupted because continuous eviction of each other's data is restricted. In this way, a spy cannot observe the cache access latency pattern to distinguish bit $0$ and $1$, resulting in noisy covert channel communication. The value of the threshold depends on the underlying replacement policy. In the rest of this paper, we use LRU to implement TPPD, and an appropriate threshold value to inhibit attack is also discussed.

\subsubsection{Structure of LLC using TPPD}
\label{sec:structure}
Figure \ref{fig:overallViewTPPD}(b) shows the structure of the LLC using TPPD. The additional components required for TPPD are shown in the two sides of the set-associative LLC. These additional components are divided into two major parts: the CCA Detector and TPPD Components. For TPPD, each set maintains a tuple, $(attack\_flag, pS, pT, CpS, CpT)$. Here the attribute, $attack\_flag$ is a single bit of information that indicates whether the set is under attack or not. The other attributes are already defined in this section. The attributes excluding $attack\_flag$, are only required for a suspicious set. When the CCA detector detects a CCA attack, the corresponding attributes of the attacker set are updated in the TPPD component. As mentioned in Section \ref{sec:detection} the CCA detection module is a cross-process conflict-miss pattern-based detector as proposed in \cite{prodact}.

\subsubsection{Engagement and disengagement with dual victim policy}
Since the CCA attack happens during execution, in the beginning, all the sets are non-suspicious. A set engages with dual victim policy only when it has been detected as targeted set by the CCA detector. Once a process pair and a set $s'$ has been detected as suspicious, one option is to continue with the proposed dual victim policy (in $s'$) till the termination of these processes. However, the option may reduce the performance of these processes in false-positive cases. The second option is to periodically check the set status in the CCA detector. If any time the cross-process misses between spy and trojan reduces in $s'$, the set may again reset as non-suspicious. Though the chances of such false-positive cases in the CCA detector are very few, we have only used the first option in this paper.

\subsubsection{Maintaining Process ID}
\label{sec:prcess-id}
The replacement policy of TPPD needs the process id of a block stored in the LLC. Figure \ref{fig:overallViewTPPD}(b) shows it as a metadata of each LLC block. The existing partitioning techniques \cite{ucp,partition-halwe} as well as some replacement policies \cite{hawkeye,prathamesh,warrier2013application} need process id for each block. Hence the overhead of maintaining process id in LLC as metadata cannot be considered as the overhead of TPPD. In case we assume only one process can run in a core, then instead of process id, we can also use the core id. Storing core id takes fewer bits than storing process id. This policy is used in some existing works \cite{ucp,partition-halwe}. However, in a practical design, each block must store the process id. The additional components required for TPPD (as shown in Figure \ref{fig:overallViewTPPD}(b)) also need to store process id. We have considered all the additional components as an overhead of TPPD. In Section \ref{sec:hardware}, we have calculated the hardware overhead of TPPD, both assuming core id and process id as TPPD components. 

\shadowbox{%
	\begin{minipage}[]{.44\textwidth}
		\begin{center}
			\textbf{Terminology used in Algorithms:} \\
			-----------------------------------------
		\end{center}
		\begin{itemize}[leftmargin=*]
			\item \textbf{$Eviction Victim(i,p)$:} Finds the the victim block to be replaced from set $i$ .
			\begin{itemize}
				\item $i$: Set number of the victim block.
				\item $p$: Owner process of the incoming block. Or the prcoess responsible for triggering the cache replacement. 
			\end{itemize}
			
			\item \textbf{$isUnderAttack(i)$: }Returns true if the set $i$ is under CCA attack, otherwise false. The value is provided by status bit maintained in additional TPPD structure.
			
			\item \textbf{$getTrojan(i), getSpy(i)$:} Returns the suspicious process pair(spy and trojan) on set $i$. The values are provided by the two suspicious process identifier maintained in TPPD structure.
			
			\item \textbf{$getLRUVictim(i)$:} Returns $V(i)$ from the set $i$. Or returning the victim block as per LRU replacement policy.
			
			\item \textbf{$checkOwner(i, w, p)$:} Returns true if process $p$ is the owner of the block $w$ in set $i$, otherwise false. 
			
			\item \textbf{$Owner(w)$:} Returns owner process of block $w$. Also represented as $O(w)$. 
			
			\item \textbf{$pS[i]$ and $pT[i]$:} Process identifier of spy and trojan respectively.
		   			
			\item \textbf{$CpS[i]$ and $CpT[i]$:} The counters to keep track of Spy and Trojan's blocks respectively in a suspicious set $i$. This is also defined in Section \ref{sec:TPPD}.
			
			\item \textbf{$updateCounter(i,p_{in},p_{out})$:} Updates $CpS(i)$ or $CpT(i)$ for the set $i$. Here $w$ is the incoming block. Here $p_{in}$ and $p_{out}$  are process id of incoming block and victim block respectively.
				
			\item \textbf{$findVictimExcept(i,p)$):} Returns $V_x(i,p)$, as discussed in Algorithm \ref{alg:alternateVictim}.

			\item \textbf{$th_{s}$ and $th_{t}$:} Threshold partition size of spy and trojan respectively. 
		\end{itemize}
	\end{minipage}
}

\begin{algorithm}
	\caption{Modified Eviction Policy for LRU}
	\label{alg:alternateVictim}
	\SetKwInput{KwInput}{Input}                
	\SetKwInput{KwOutput}{Output}              
	\DontPrintSemicolon
	
	\KwInput{$sId$: Cache set index. $omitP$: Process whose block not to evict.}
	\KwOutput{A victim block not belonging to process $omitP$.}
	\SetKwFunction{FSum}{findVictimExcept}
	\SetKwProg{Fn}{Function}{:}{end}
	\Fn{\FSum{$sId$, $omitP$}}{ 
		
		{/*Assuming $cache[row,assoc]$ is the set-associative LLC where $row$ is the total sets and $assoc$ is the associativity. $victim\_index$ and $max\_age$ are two variables.*/}\;
		$k=0$\;
		\While{$k \neq assoc$}{
			\If{$checkOwner(sId,cache[sID,k],omitP)$ is FALSE}{
				$victim\_index=k$\;
				$max\_age=cache[sID,k].age$\;
				\textbf{break}
			}
			k++
		}
		\While{$k \neq assoc$}{
			\If{${(max\_age < cache[sId, k].age)}$}{ 
				\If{$checkOwner(sId,cache[sID,k],omitP)$ is FALSE}{
					$victim\_index=k$\;
					$max\_age=cache[sID,k].age$\;
				}
			}
			k++
		}
		\KwRet $cache[sId,victim\_index]$\;
	}
\end{algorithm}

\begin{algorithm}
	\caption{Modified LRU Replacement Policy for implementing TPPD}
	\label{Algo:TPPDAlgo}
	\SetKwInput{KwInput}{Input}                
	\SetKwInput{KwOutput}{Output}              
	\DontPrintSemicolon
	
	\KwInput{$i$: set index, $p$: process requesting replacement.}
	\KwOutput{$w$: way index of block to be evicted from set $i$ }
	
	\SetKwFunction{FMain}{EvictionVictim}
	\SetKwComment{Comment}{/* }{*/}
	\SetKwFunction{FSub}{updateCounter}
	
	\SetKwProg{Fn}{Function}{:}{end}
	
	\Fn{\FMain{$i$, $p$}}{
		$w$=$getLRUVictim(i)$\;
		\eIf{$isUnderAttack(i)$ is FALSE}{
			\KwRet $w$ \tcp*[l]{Return $V(i)$.}
		}{\Comment*[l]{The set $i$ is under attack.}                                        
		$pT$=$getTrojan(i)$; 
		$pS$=$getSpy(i)$\;
		\uIf{$checkOwner(i,w,pS)$ is TRUE}{
			\tcp*[l]{If the owner of $w$ is spy.} 
			$p_w=pS$\;
		}\uElseIf{$checkOwner(i,w,pT)$ is TRUE}{
		\tcp*[l]{If owner of $w$ is trojan.}
		$p_w=pT$\;
	}\Else{ 
	$p_w=-1$ \tcp*[l]{if owner of $w$ is an innocent process.}
}

\uIf{($(p!=pS)\&(p!=pT))) || (p==p_w)$}{ \Comment*[l]{Incoming block from innocent process or both incoming and victim from same process.} 
	$updateCounter(i,p,p_w)$; 
	\KwRet $w$\;
}\uElseIf{$((p_w==pS)\And (CpS[i]<th_s))||((p_w==pT)\And (CpT[i]<th_t))$}{ 
$w''=findVictimExcept(i,p_w)$\; 
\If{$checkOwner(i,w',p)$ is FALSE}{
	$updateCounter(i,p,-1$)\;
}
\KwRet $w''$
}
\Else{
	$updateCounter(i,p,p_w)$;
	\KwRet $w$
}

}
}

\SetKwProg{Fn}{Def}{:}{end}
\Fn{\FSub{$i$, $p_{in}$, $p_{out}$}}{ \Comment*[l]{$p_{in}$ is the process of incoming block and $p_{out}$ is the process of outgoing (or victim) block.}
	\If{$p_{in} != p_{out}$}{
		\uIf{$p_{out}==pS$}{
			$CpS[i]--$ \; 
		}\uElseIf{$p_{out}==pS$}{
		$CpT[i]--$ \; 
	}\textbf{end}
	
	\uIf{$p_{in}==pS$}{
		$CpS[i]++$ \; 
	}\uElseIf{$p_{in}==pS$}{
	$CpT[i]++$ \;
}\textbf{end}
}
}
\end{algorithm}

\subsection{TPPD for LRU Replacement Policy}
Algorithm \ref{alg:alternateVictim} shows an implementation of $V_x(s',p)$ for LRU replacement policy. The alternative victim ($V_x(s',p)$) in LRU can be selected in two ways. The first method is selecting a random block not belonging to $p$, and the second method is to choose the oldest block that does not belong to $p$. The algorithm describes the second method. All the experimental analyses in this paper use the second method. The complete mechanism of TPPD in terms of LRU replacement policy is discussed in Algorithm \ref{Algo:TPPDAlgo}. The terminology used in this algorithm is described in the box. As mentioned in Section \ref{sec:TPPD}, we have used $s'$ to represent the targeted set, but to write an algorithm, we have used $i$ to represent any LLC set. The set can be either suspicious or non-suspicious. Here $0 \le i < N$, and $N$ is the total sets of LLC.

The main function in this algorithm is $Eviction Victim(i,p)$, which selects the appropriate victim block from the LLC as per the requirement of TPPD. The function is called every time a process $p$ requests for a block and has a conflict miss. Process $p$ is called an incoming process, and the block it requested as an incoming block. Initially, an original eviction victim $w$ is selected per the LRU replacement policy (Line 2). Then it checks if the requested set is under attack or not (Line 3). If not, the original victim block $w$ is returned (Line 4). However, additional checks are required if the request is for the suspicious set. Everything mentioned from Line 5 onward in this function is for a suspicious set. First, the owner process $p_w$ of the victim block $w$ is determined (Line 6 to Line 13). The $p_w$ can be either spy ($pS$), trojan ($pT$), or an innocent process. If the incoming process $p$ is innocent, it can replace any block from the set. Hence, in that case, $w$ is evicted irrespective of $p_w$ and the corresponding counter $(CsT[i]$ or $CpT[i]$ based on $p_w)$  needs to be updated. Line 14 and 15 do the same. The function $updateCounter(i, p_{in}, p_{out})$ is used for this which is discussed later. Similarly, the original victim $w$ is evicted when the incoming block and victim block belong to the same process but without an update in counters.

The condition mentioned in Line 16 is true only when these three conditions are true: (a) $p$ and $p_w$ are not the same, (b) the incoming process $p$ is either trojan or spy, and (c) the value of counter corresponding to $p_w$ is less than the threshold. In this case the alternative victim block $w''$ is selected by calling function $findVictimExcept(i,p)$ (Line 17). This function is defined in Algorithm \ref{alg:alternateVictim}. New victim block $w''$ can be from the incoming process $p$ or from an innocent process. The counter update is required only in the latter case (Line 18-20). In case the condition written in Line 16 is false then the original LRU victim $w$ is selected as victim block (Line 23), and counters are updated. Function $updateCounter(i, p_{in}, p_{out})$ keeps spy and trojan updated based on the id of incoming process $(p_{in})$ and process of victim block $( p_{out})$.  Note that no counter updates are required if $p_{in}$ and $p_{out}$ are the same. Also, counter updates are not required in non-suspicious sets. Since the counter maintains the total number of blocks available in the set for trojan and spy, they may need to be incremented or decremented during an eviction. From the algorithm, it can be observed that TPPD applies dual victim policy only to the suspicious sets. Also, the need for selecting the alternative victim is only required when trojan and spy tries to remove each other's block. Section \ref{sec:exp} shows experimental analysis on different threshold values $(th_{s}$ and $th_{t})$ that can be used for this algorithm.

\subsection{How does TPPD effectively mitigate covert channel attack?}
\label{sec:howmitigate}
In this part, we give a theoretical analysis of how TPPD dismantles LLC-based cross-core covert channel attacks with minimal impact on system performance. Our observations and their explanations are described as follows:

\textbf{Observation 1:} 
The communication between the spy and the trojan of the covert channel attack is degraded when the suspicious process pair is prevented from accessing all the ways of a targeted cache set.

\textbf{Explanation: }
A low and high cache access latency pattern observed by a spy to determine the bit transmitted by trojan is at the foundation of a successful covert channel attack. Trojan establishes this pattern by removing all of the spy's data in one case (e.g., bit $1$) or none at all for another bit (bit $0$). In TPPD design, cache interference between spy and trojan is decreased by restricting them from evicting each other' block when the current number of blocks of these processes falls below the threshold. As a result, the trojan is unable to replace all of the spy's data for sending bit 1, and the spy cannot fully prime the cache set by replacing trojan's data with its. In this way, the spy's easily identifiable access latency pattern is obscured.

\textbf{Observation 2: }A targeted defence mechanism that affects only suspected sets and processes limit the overall system performance degradation.

\begin{figure*}[h]
	\centering
	\includegraphics[scale=0.65]{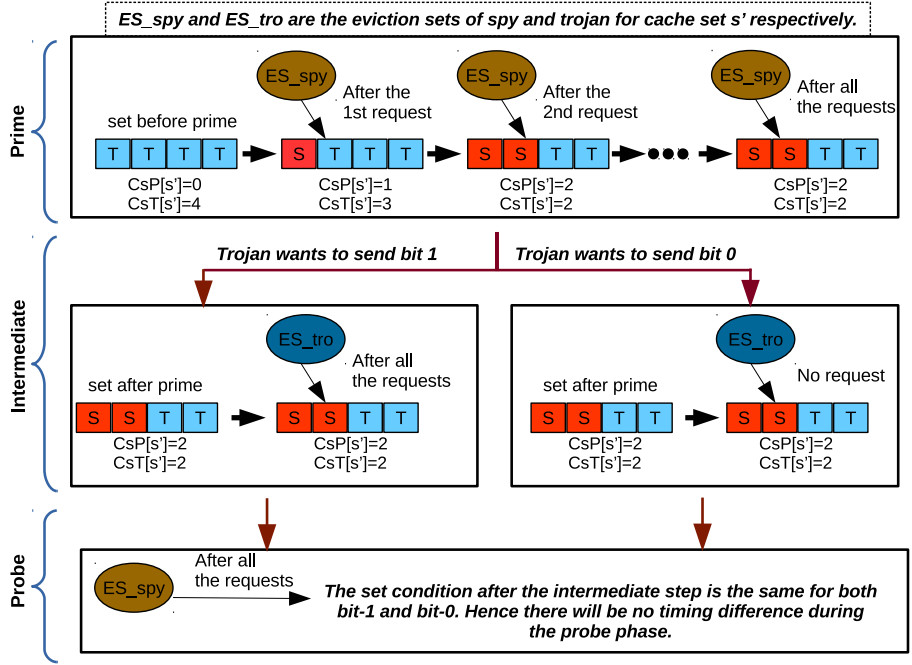}
	\caption{An example to demonstrate the working of TPPD after detecting a CCA on set $s'$ for $pS$ (as spy) and $pT$ (as trojan). We have assumed that the set $s'$ contains all the blocks of trojan during the detection. The prime process starts with this assumption. The TPPD configuration considered here is TPPD-$2$. The figure shows that the Prime+Probe method cannot perform CCA once TPPD-$2$ becomes active. }
	\label{fig:exampleTPPD}
\end{figure*}

\textbf{Explanation}: 
The main feature of our proposed defense mechanism is that it is a targeted countermeasure. It affects only processes and cache sets detected to be participating in covert channel attack by the CCA detector. Modified replacement policy to implement TPPD design is activated on suspicious sets, and it restricts only conflict misses occurring between the suspicious processes. Original replacement policy is implemented for remaining sets not participating in covert channel attack. On suspicious sets, trojan and spy processes have restrictions on evicting each other's block. The innocent processes have access to the entire cache set and can evict block of any process as determined by the original replacement policy. However, when suspicious processes are restricted from evicting each other's block, a new block is selected for eviction. This can lead to premature eviction of benign process blocks, thus slightly impacting overall system performance.

\subsection{Calculating Performance of TPPD}
TPPD can use different threshold values for $th_s$ and $th_t$. These values can be either the same or different for spy and trojan. In the rest of the discussion, we have considered the same values ($th_s = th_t$) for both spy and trojan. The TPPD-$z$ is considered as a configuration having $th_s = th_t = z$. The minimum value of $z$ is 1, and the maximum is $A/2$, where $A$ is the associativity of the LLC. The value of $z$ can be changed dynamically, but for most of our experiments, we have fixed the value of $z$ throughout the execution. 

We have measured the isolated performance impact on benign applications through the following steps:
\begin{itemize}
	\item When only the attacker processes are executing in the system, the impact of TPPD-$z$ in terms of LLC misses ($DIFF_{z}$) is calculated by differentiating number of misses in LLC having TPPD-$z$ and  no defense.
	\begin{equation}
	\label{eqn:diff}
	DIFF_{z} = Misses_z - Misses_0 \hspace{8pt} 
	\forall z\in (1, A/2) 
	\end{equation}
	Here $Misses_{z}$ means total misses on LLC while TPPD-$z$ is active. $Misses_{0}$ means total LLC misses when no defense mechanism is implemented in LLC.
	
	\item In the next scenario, a Parsec benchmark \cite{parsec2009} is run parallel to the attack processes. The difference in LLC misses ($DIFF'_{z,b}$) encountered in TPPD-$z$ ($Misses_{z,b}$) and without defense ($Misses_{0,b}$ ) is measured for benchmark $b$. 
	
	\begin{equation}
	\label{eqn:diff'}
	DIFF'_{z,b} = Misses_{z,b} - Misses_{0,b} \hspace{8pt} 
	\forall b\in PARSEC 
	\end{equation}
	Here $Misses_{z,b}$ means total LLC misses while TPPD-$z$ is active. $Misses_{0,b}$ means total LLC misses without a defence mechanism. 
	
	\item The isolated impact of TPPD-$z$ on a benign application ($D_{z,b}$) and average impact($AvgD_{z}$) is calculated as described in Equation \ref{eqn:d} and \ref{eqn:AvgD} respectively.
	
	\begin{equation}
	\label{eqn:d}
	D_{z,b} =  DIFF'_{z,b} -  DIFF_{z}
	\end{equation}
	
	\begin{equation}
	\label{eqn:AvgD}
	AvgD_{z}=\sum_{b \in PARSEC } \frac{D_{z,b}}{T_n}
	\end{equation}
	Here, $T_n$ is the total number of PARSEC applications considered. 
\end{itemize} 
These equations are used in Section \ref{sec:exp} for performance analysis. The higher positive difference ($AvgD_{z}$) indicates more significant performance degradation of benign processes. A difference close to zero will signify that there has been no effect on these processes.

\subsection{An example of TPPD}
An example discussing how our proposed approach deteriorates covert channel communication on LLC is shown in Figure \ref{fig:exampleTPPD}. Consider the Prime+Probe based CCA attack mounted on set $s'$ of LLC. After attack detector detects this attack; it sends the set id $s'$ along with process ids $pS$ and $pT$ to TPPD. When TPPD-$2$ is deployed initially, we assumed the entire cache set $s'$ contains trojan's block (trojan is sending 1) as shown in Step 2 of Figure \ref{fig:ppexample}. This assumption is considered without any loss of generality and a better understanding of TPPD working. 

During the prime phase, the spy requests for blocks from its eviction set. From the figure, it can be observed that at the end of all the block requests of the prime phase, the spy could not be able to remove all the trojan's block from the set. This is because of the dual victim policy of TPPD-$2$. In its first two accesses, the spy is able to replace trojan's block as intended. However, for the next spy access, if the original LRU victim spy block is evicted, the number of spy's blocks present in the set falls below the threshold of 2, which is not allowed by TPPD-$2$. In this case, as per the dual victim policy of TPPD, an alternative replacement victim that does not belong to the trojan process is selected using $V_x(s', pT)$. After the prime phase, in the case of sending bit 1, the trojan removes all the blocks of spy by requesting block from its own eviction set. However, because of TPPD-$2$, the trojan cannot remove all the spy's blocks. From the figure, it can be observed that, after the intermediate phase, the status of the set is the same for both bit 1 and bit 0. In the probe phase, when spy probes set $s'$ for the blocks from its eviction set, it encounters $2$ misses and $2$ hits in both cases of a bit $1$ and bit $0$.

\subsection{Implementing TPPD on other Replacement Policies}
The proposed TPPD can be implemented using any replacement policy where dual victim policy can be implemented. Algorithm \ref{Algo:TPPDAlgo} can be modified for that. The internal mechanism of the functions called from Line 2 and Line 21 of this algorithm needs to be changed as per the requirement of the replacement policy. In this work, we have not explored the TPPD for any other replacement policy except LRU, but that can be explored without any major changes. The proposed TPPD can be considered as a countermeasure for cross-core CCA that can make any existing replacement policy secure from CCA.

\begin{table}[]
	\centering
	\caption{System configuration for the experimental setup}
	\begin{tabular}{|p{2.5cm}|p{3.5cm}|}
		\hline
		\multicolumn{2}{|c|}{\textbf{Baseline Architecture}}                                           \\ \hline
		Simulator                   & gem5 \cite{gem5}                                      \\ \hline
		Architecture                & 4 cores each at 2.0 GHz                                              \\ \hline
		Coherence Protocol          & MESI two level                                                   \\ \hline
		Level 1 Cache               & Inclusive L1I/L1D Private, 64 KB, 4-way, 2 cycles latency, 64B blocks                \\ \hline
		Level 2 Cache (LLC)         & Inclusive Shared, 2MB 8-way, 18 cycles latency, 64B blocks                         \\ \hline
		Main Memory                 & DRAM, 250-cycle latency \\ \hline
		Replacement Policy &     Least Recently Used(LRU)                                                             \\ \hline
	\end{tabular}
	\label{tab:parameters}
\end{table}

\begin{figure}[]
	\centering
	\includegraphics[scale=0.45]{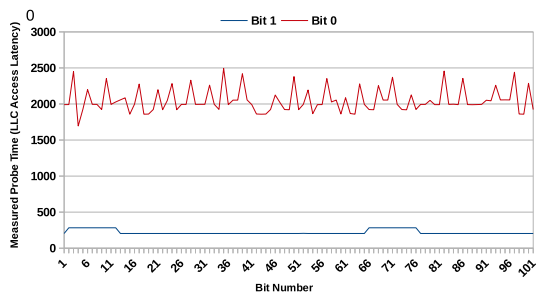} 
	\caption{Cache access latency observed by spy for bit $0$ and $1$ in LLC with no TPPD}
	\label{fig:time00}
\end{figure}

\begin{figure*}[]
	\begin{subfigure}{.5\textwidth}
		\centering
		
		\includegraphics[scale=0.45]{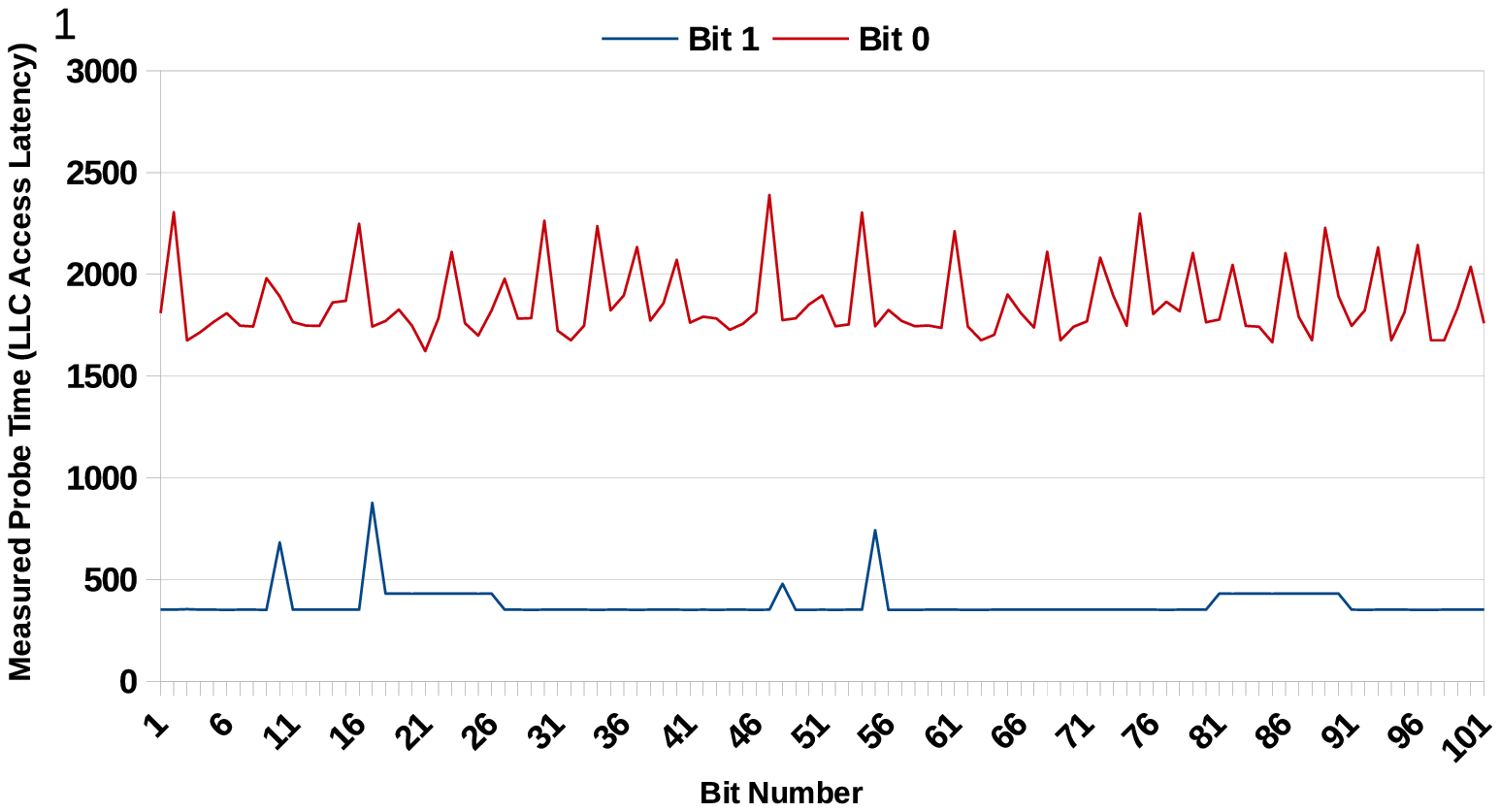}  
		\caption{}
		\label{subfig:time11}
	\end{subfigure}
	\begin{subfigure}{.5\textwidth}
		\centering
		\includegraphics[scale=0.45]{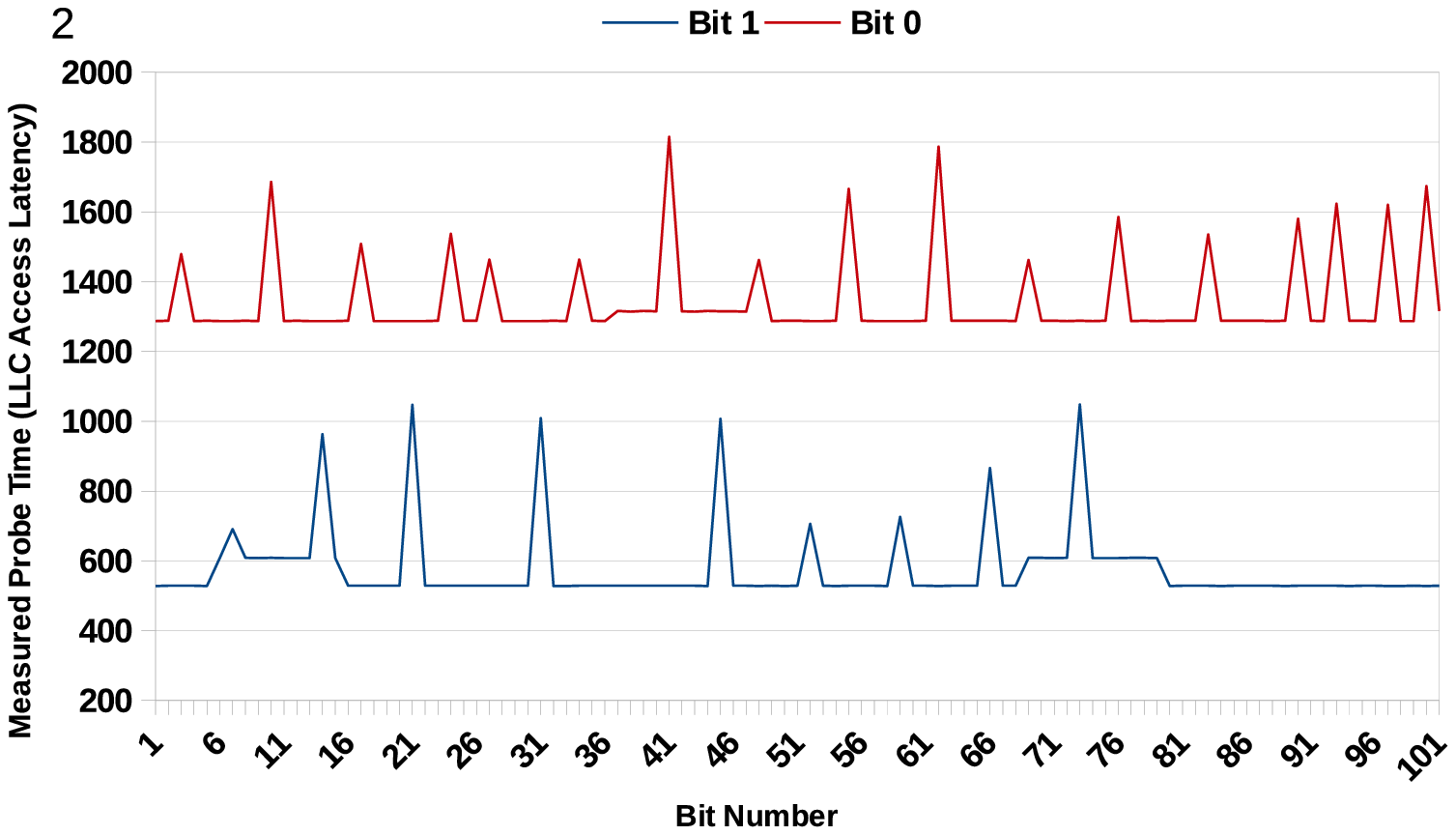}  
		\caption{}
		\label{subfig:time22}
	\end{subfigure}
	
	\begin{subfigure}{.5\textwidth}
		\centering
		\includegraphics[scale=0.45]{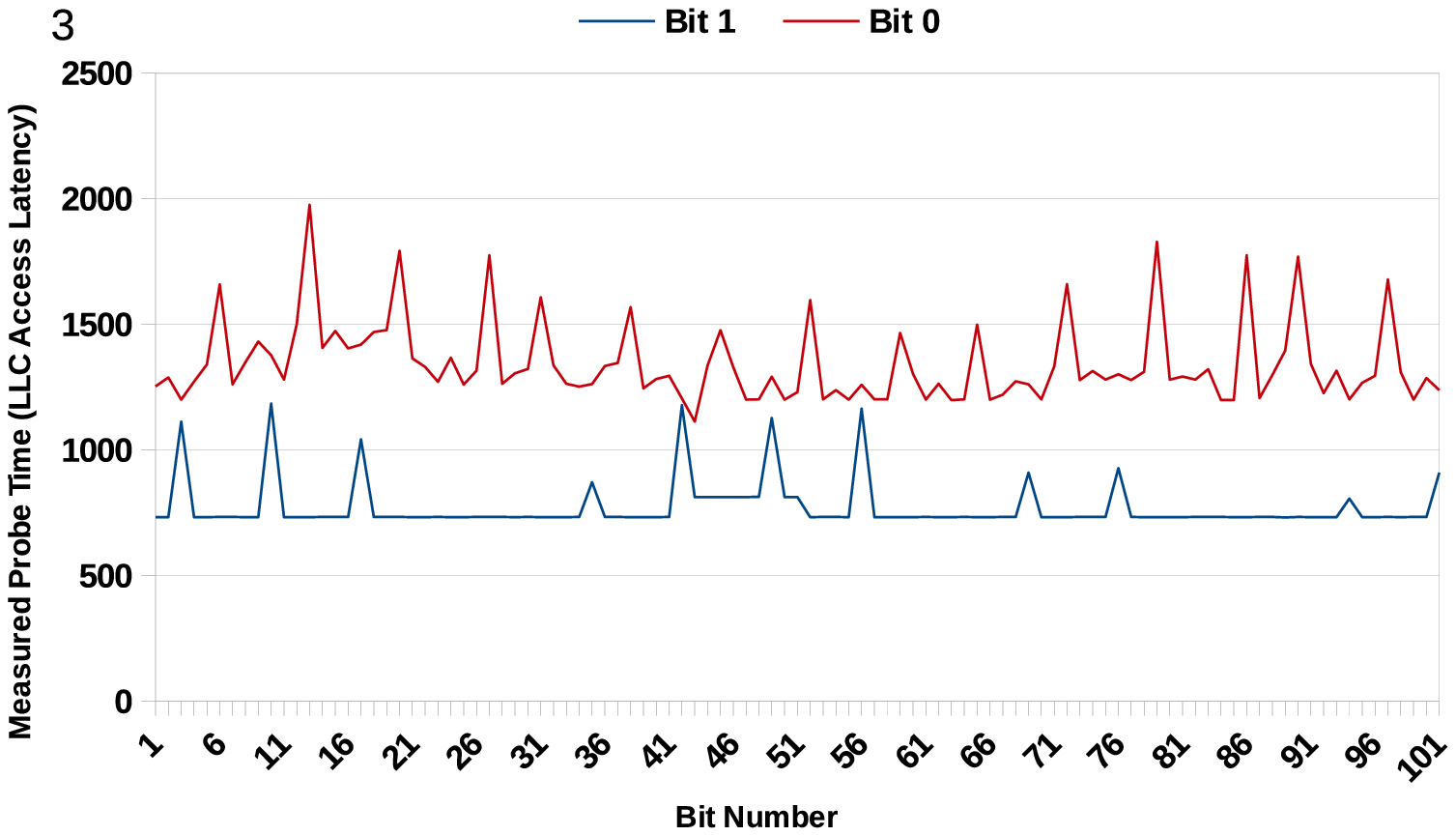}  
		\caption{}
		\label{fig:time33}
	\end{subfigure}
	\begin{subfigure}{.5\textwidth}
		\centering
		\includegraphics[scale=0.45]{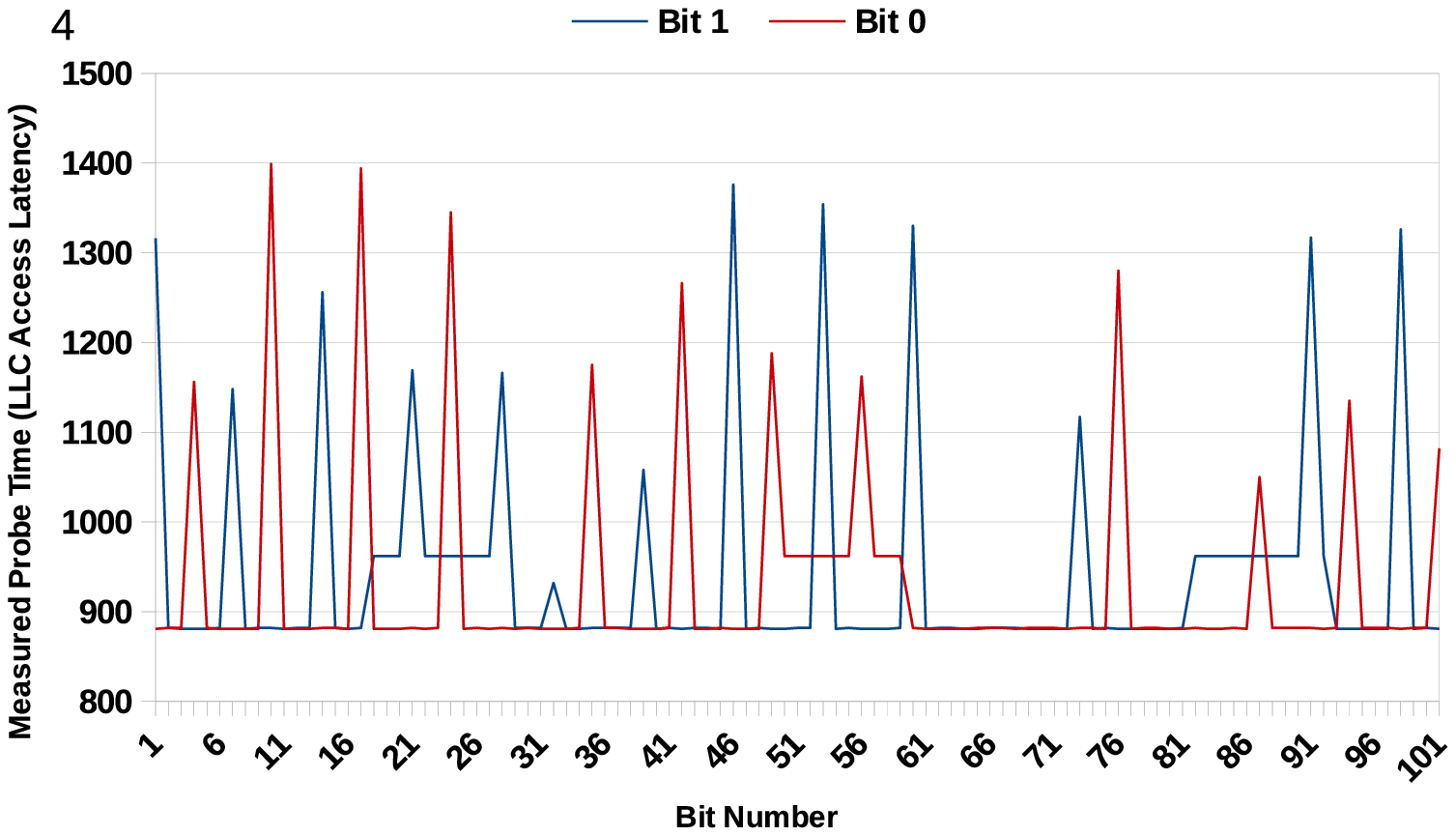}  
		\caption{}
		\label{subfig:time44}
	\end{subfigure}
	\caption{Cache access latency for bit $0$ and bit $1$ in (a) TPPD-1, (b) TPPD-2, (c) TPPD-3, and (d) TPPD-4}
	\label{fig:timeTPPD}
\end{figure*} 

\section{Experiments and Results}
\label{sec:exp}
To perform the experimental analysis, TPPD is implemented on a full system cycle-accurate simulator, gem5 \cite{gem5}. We consider a 4-core system setup with two levels of the cache hierarchy. Each core has its private L1 cache, and the L2 cache is considered a shared LLC. Other parameters of the setup are shown in Table \ref{tab:parameters}. To simulate the CCA attack, we have developed our own trojan and spy applications. These applications are written in C++ and can be executed on gem5. We considered a generic Prime+Probe attack as described in Figure \ref{fig:ppexample} in this application. The ability to perform CCA attack by these two applications have experimented first on gem5. Parsec benchmarks \cite{parsec} are used to measure the performance of the innocent applications in the presence of attacker applications. From the 4-cores, two cores are assigned to the attacker processes (spy and trojan). A multi-threaded Parsec benchmark is binded to the other two cores. We have considered all the Parsec benchmarks as innocent applications. To measure the worst-case performance impact, the attacking applications are developed such that once the attack starts, it continues during the execution of the system. Different configurations of TPPD (TPPD-$z$) are considered in order to fully analyse the performance behaviour of TPPD. The value of $z$ varies from $1$ to $4$ ($8$ is the associativity of the LLC). TPPD-$0$ means baseline design with no defence mechanism.

\subsection{Security Analysis}
We assess the effectiveness of TPPD against cross-core based covert channel attacks on LLC. As described in Section \ref{sec:howmitigate}, TPPD reduces the difference in probe time observed by the spy for bit $0$ and $1$. This difference is substantially more significant when no defense mechanism is active, as evident by Figure \ref{fig:time00}. Here bit $0$ represents cache miss and bit $1$ as a cache hit. TPPD reduces this difference such that the information received by the spy becomes noisy. This difference becomes close in TPPD-1, TPPD-2 and TPPD-3 but does not overlap as represented in Figure ~\ref{subfig:time11}, \ref{subfig:time22} and \ref{fig:time33} respectively. These TPPD configurations are effective against attacks based on fine-grained information, i.e., observing the number of cache accesses by victim applications in the targeted set. However, in a low-speed covert channel attack, we considered these configurations would not be effective, especially when there is low system noise. However, TPPD-4 was able to completely obfuscate the receiver's access time pattern used to identify bit $0$ and $1$ as illustrated in Figure \ref{subfig:time44}. Hence, TPPD-$A/2$ is the recommended configuration to prevent most CCA attacks. Here $A$ is the associativity of the LLC.

\begin{figure*}[]
	\centering
	\includegraphics[scale=0.6]{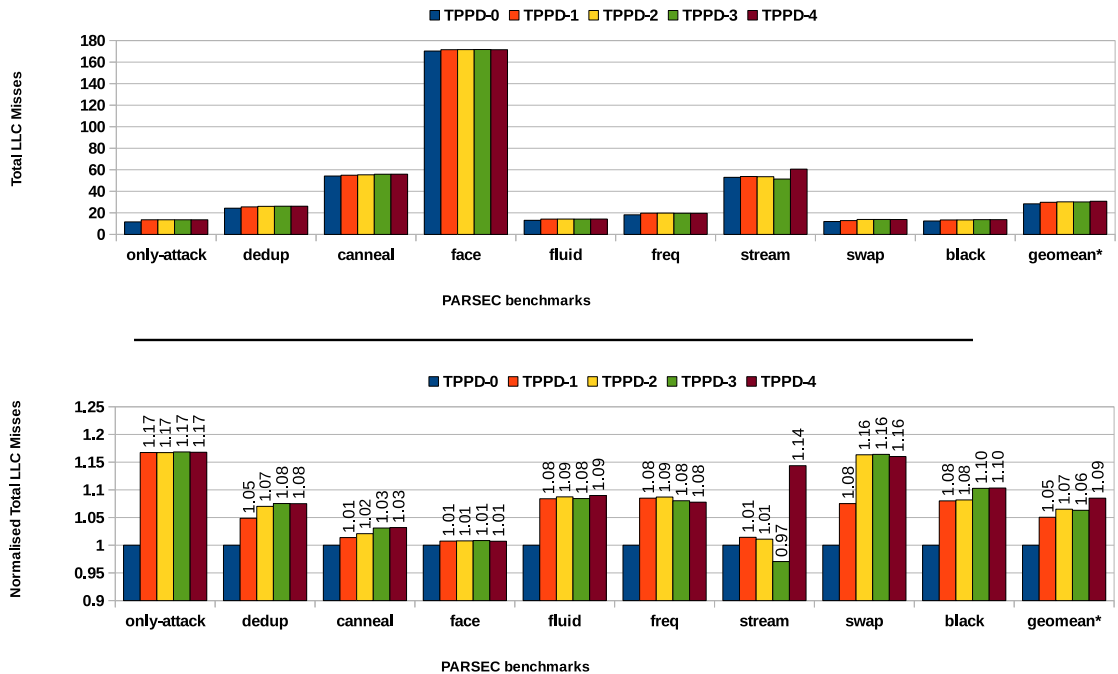}
	\caption{Total number of misses in LLC for PARSEC benchmarks on different TPPD configurations. *The calculation of geomean is excluding ``only-attack''. The lower part of the figure shows the LLC misses normalised to TPPD-0.}
	\label{fig:totalllcmisses}
\end{figure*}
\begin{figure*}
	\centering
	\includegraphics[scale=0.6]{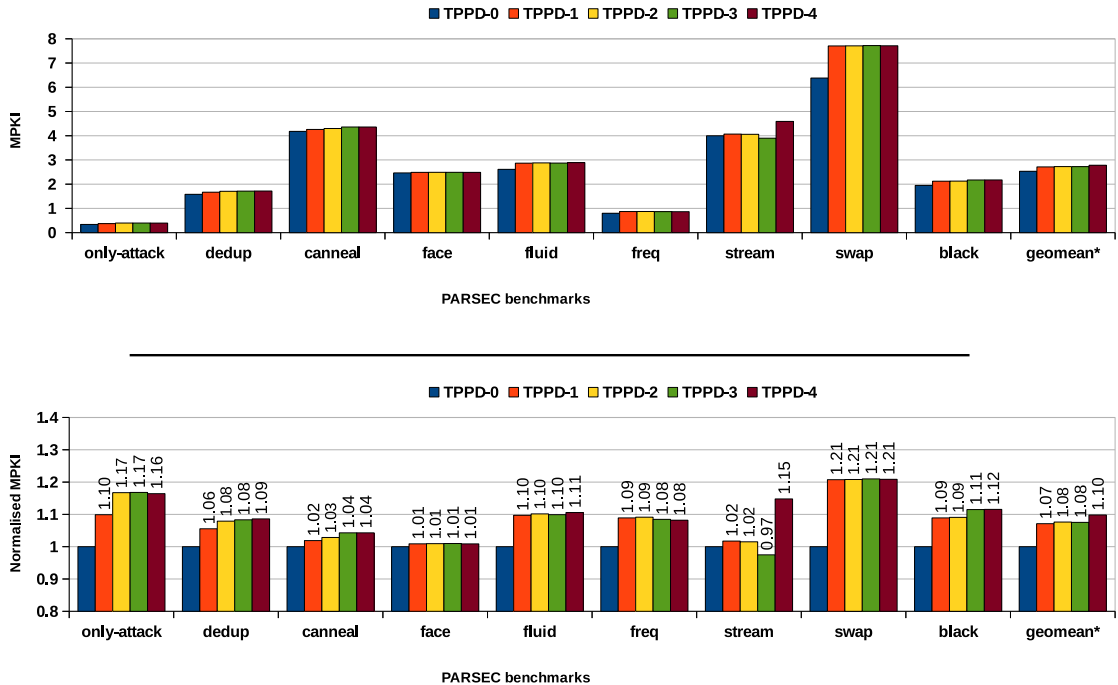}
	\caption{Total MPKI in LLC for PARSEC benchmarks on different TPPD configurations. *The calculation of geomean is excluding ``only-attack''. The lower part of the figure shows the MPKI normalised to TPPD-0.}
	\label{fig:mpki}
\end{figure*}

\begin{figure}[]
	\includegraphics[scale=0.45]{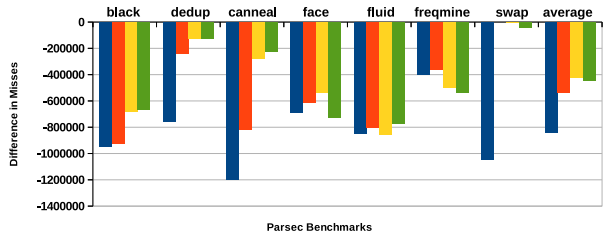} 
	\caption{Isolated LLC misses for PARSEC benchmarks with different TPPD configurations measured using Equation \ref{eqn:d}}
	\label{fig:differenceInMisses}
\end{figure}

\begin{figure}[]
	\includegraphics[scale=0.45]{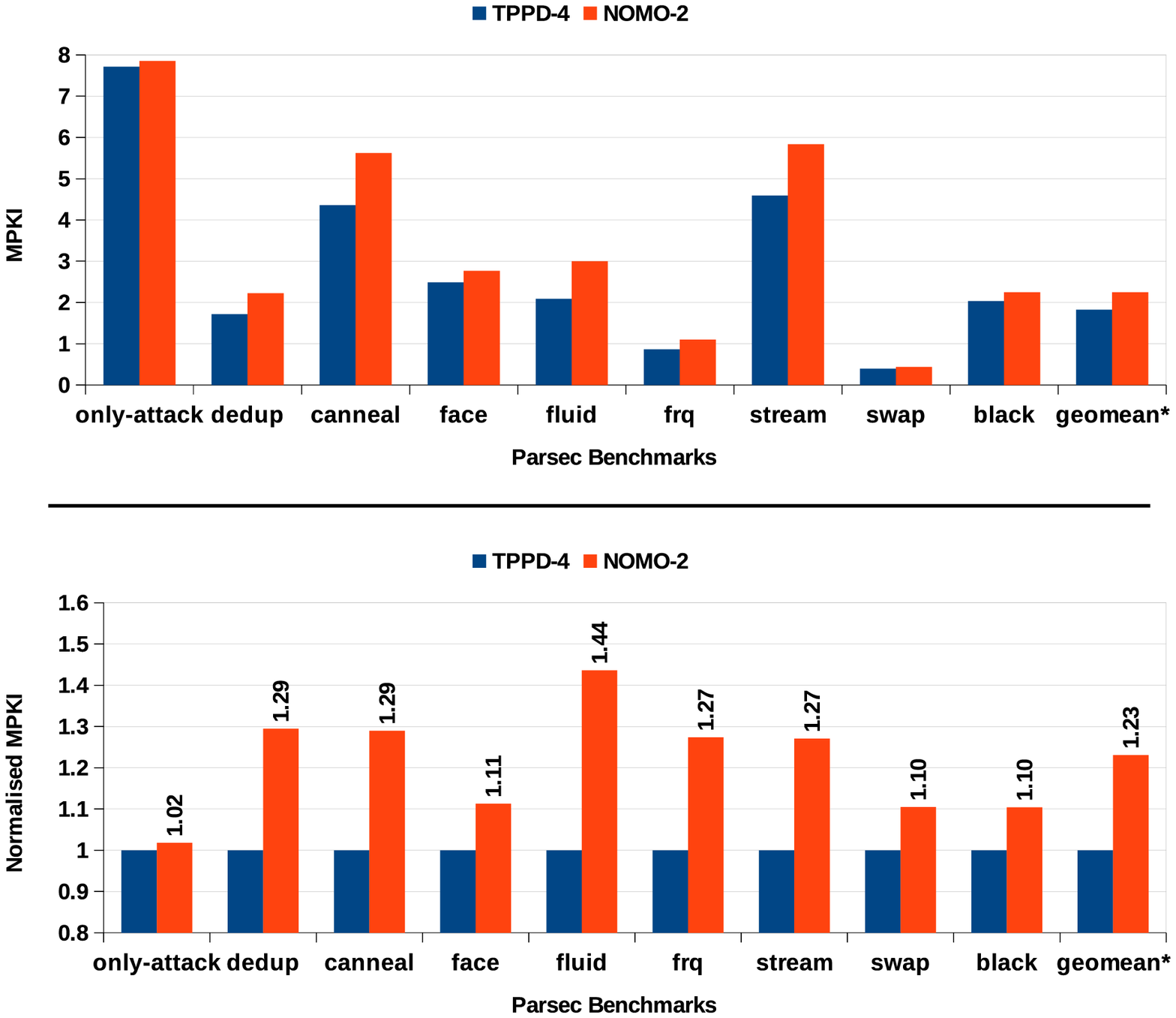} 
	\caption{Comparison of MPKI for NOMO-2 vs TPPD-4 when PARSEC benchmark are executed in parallel to CCA. Here *geomean is calculated excluding ``only-attack''.}
	\label{fig:NOMO2vsTPPD4}
\end{figure}

\subsection{Performance Analysis}
While deconstructing cache timing channel-based attacks, the attack defense mechanism must guarantee that system performance does not suffer. The Figure \ref{fig:totalllcmisses} represent the total LLC misses encountered in different benchmark applications for TPPD-$z$ and TPPD-$0$. The upper part of the figure shows the actual values (in millions), while the bottom part shows the normalised LLC misses. The values are normalised w.r.t. TPPD-0. As mentioned above, the benchmarks are combined with the attacker applications during the execution. However, in the figure, the benchmark named as "only-attack" is not a Parsec benchmark. This is designed only with the combination of trojan and spy. To analyse the performance impact of TPPD on innocent applications, it is important to first analyse the performance of "only-attack". From the figure, it can be observed that, in the case of "only-attack", if TPPD is applied to prevent CCA, the LLC misses are increased by 17\% as compared to TPPD-0 (no defence). All the configurations of TPPD show almost similar results. However, for the rest of the benchmarks, the increment is less. On average, while a Parsec benchmark executes with two attacker applications, which are constantly trying to covertly communicate, the performance degradation in TPPD are 05\%, 07\%, 06\% and 09\% for TPP-1, TPPD-2, TPPD-3, and TPPD-4 respectively. The degradation in TPPD-4 is more because a higher threshold value may evict some more innocent blocks from the cache. A similar result can be seen in Figure \ref{fig:mpki} for Misses per Thousand Instructions (MPKI).

The important point here is that the misses are higher when TPPD applies only to the attacker set in "only-attack". However, the miss overhead is reduced when the cores mix attack applications with some innocent Parsec benchmarks. It means that the miss overhead of TPPD is higher only for the attacker applications. The overhead is less for innocent applications. The two main reasons for this are:
\begin{itemize}
	\item TPPD is a targeted CCA countermeasure. Hence from the non-targeted set, the victim is selected per the original replacement policy.
	\item In the targeted set, the innocent blocks has no restrictions, and they can remove any other block from the set.
\end{itemize}
Because of these two reasons, TPPD becomes an efficient countermeasure for cross-core covert channel attacks.

Figure \ref{fig:totalllcmisses} and \ref{fig:mpki} only shows the combined miss overhead when the innocent applications are running with attacker applications. To understand the isolated impact of TPPD on innocent applications, we have performed experiments based on Equation \ref{eqn:d} and Equation \ref{eqn:AvgD}. As evident from Figure \ref{fig:differenceInMisses}, the increment in total LLC misses (compared to TPPD-0) when benchmark runs alongside attack process is always less than the misses when just attack runs in LLC augmented with TPPD-$z$. In this figure, a less number (large negative number) means less overhead on an application. On average, the applications face the lowest overhead in TPPD-1. However, the miss overhead in TPPD-4 (which is the recommended configuration) is also less.

\subsection{Comparison with Existing Partitioning based Secure Cache Architectures}
Existing partitioning-based approaches such as PLCache\cite{newcachedesigns}, SecVerilog Cache\cite{secverilog}, and SecDCP\cite{secdcp} can combat cache-based side-channel attacks in which an attacker process attempts to reveal the cache access pattern of an innocent victim process in order to leak its secret, such as the private key of cryptography algorithms. These secure cache designs rely on a reliable operating system, compiler, or programmer to recognise security-critical processes or data that could be vulnerable to cache timing channel attacks. PLCache pre-loads and lock the security critical data in the cache and prevents its eviction by unlocked data. In SecVerilog and SecDCP, processes are assigned to different security classes, and cache interference between these security domains is restricted. These security measures are ineffective in the presence of covert timing channel attacks where both processes are malicious.

The static cache partitioning based secure cache design: NOMO\cite{nomo} can defend against covert channel attacks by completely isolating ways of spy and trojan. In our setup, NOMO-2, in which two ways were fixed per process, was able to dismantle the covert channel attack. However, significant performance degradation was observed in the case of NOMO-2 in comparison to our design TPPD-4 as shown in Figure \ref{fig:NOMO2vsTPPD4} in terms of MPKI. It can be observed from the figure that NOMO shows less miss overhead over TPPD-4 when executed for ``only-attack''. However, when innocent applications are executed with attacker applications, the overhead increases significantly. The main reason for that is the non-targeted nature of NOMO. The restricted eviction mechanism is uniformly applied to all the sets. Hence, the overhead of misses in increases in non-targeted sets also. Since the attackers mainly requested for the blocks mapped to the targeted set, the other set mostly served innocent applications. Uniform applications of restriction eviction policy impact the miss rate of innocent application in non-targeted sets. On average, NOMO-2 had $23\%$ higher MPKI than TPPD-4.

\subsection{Storage and Latency Overhead}
\label{sec:hardware}
The structure of the LLC having TPPD as a countermeasure is already discussed in Section \ref{sec:structure}. Figure \ref{fig:overallViewTPPD}(b) shows the two additional components required for TPPD: (a) TPPD Components, and (b) CCA Detector. In this section, we have discussed the storage and latency overhead of the TPPD components. As mentioned in Section \ref{sec:detection}, we have used the existing CCA detector technique \cite{prodact} to detect the CCA attack. Each cache line is augmented with its owner process id to identify the cache block's owner process, as seen in Figure \ref{fig:overallViewTPPD}(b). An alternative technique followed by some existing works \cite{ucp,partition-halwe} is to maintain core id instead of process id along with each cache line. However, the overhead of storing process id along with the cache line cannot be considered as the overhead of TPPD. The issues with maintaining process id along with the cache lines are already discussed in Section \ref{sec:prcess-id}.


Figure \ref{fig:overallViewTPPD}(b) shows all the counters used in TPPD Components. The first counter, $attack\_flag$, is a one-bit counter used to indicate whether or not the set is under attack. The next two fields ($pS$ and $pT$) specify the attacking process pair if the set is under attack. The last two fields are spy and trojan counters, each with a size of $log_2(A)$ bits to identify the number of spy and trojan blocks currently in the set. Here, $A$ is the associativity of the LLC. Thus, the total storage overhead can be determined as $N \times(1+ 2(y+log_{2}A))$. Here $N$ is the total number of sets in LLC, and $y$ is the bits required to represent a suspicious process. In our work, $y$ represents the cores in the system as a single process is binded per core making core id enough for identifying attacking pair. Table \ref{tab:storageOverhead} shows the storage overhead of TPPD as per the LLC configurations mentioned in Table \ref{tab:parameters}. It can be observed from Table \ref{tab:storageOverhead} that the total storage required for TPPD is $5.5 KB$, i.e.  $0.26\%$ of total LLC size. However, in real systems, multiple processes may run on the same core, making core id not sufficient for uniquely identifying processes. The table shows that when process id is used instead of core id, the total size of this additional structure is $20.5 KB$, i.e. $\approx 1\%$ of total LLC size. 


The operations of TPPD are performed in the background without affecting the critical execution path of the system.

\begin{table}[]
\caption{Storage overhead calculation of TPPD as per the LLC configuration mentioned in Table \ref{tab:parameters}. We have assumed 16 bits to represent a process id as the maximum process id possible in Linus is $32768$.}
\label{tab:storageOverhead}
\centering
\begin{tabular}{|c|cc|cc|}
\hline
\multirow{2}{*}{\textbf{Storage Overhead}} &
  \multicolumn{2}{c|}{\textbf{$y$ as core id}} &
  \multicolumn{2}{c|}{\textbf{$y$ as process id}} \\ \cline{2-5} 
 
 &
  \multicolumn{1}{c|}{per set} &
  per LLC &
  \multicolumn{1}{c|}{per set} &
  per LLC \\ \hline
  
\begin{tabular}[c]{@{}c@{}}status \\ bit\end{tabular} &
  \multicolumn{1}{c|}{1} &
  512 Bytes &
  \multicolumn{1}{c|}{1} &
  512 Bytes \\ \hline
  
\begin{tabular}[c]{@{}c@{}}suspicious\\ process1\end{tabular} &
  \multicolumn{1}{c|}{2 bits} &
  \begin{tabular}[c]{@{}c@{}}2*4096\\ =1 KB\end{tabular} &
  \multicolumn{1}{c|}{16 bits} &
  \begin{tabular}[c]{@{}c@{}}16*4096\\ =8 KB\end{tabular} \\ \hline
  
\begin{tabular}[c]{@{}c@{}}suspicious\\ process2\end{tabular} &
  \multicolumn{1}{c|}{2 bits} &
  \begin{tabular}[c]{@{}c@{}}2*4096\\ =1 KB\end{tabular} &
  \multicolumn{1}{c|}{16 bits} &
  \begin{tabular}[c]{@{}c@{}}16*4096\\ =8 KB\end{tabular} \\ \hline
  
\begin{tabular}[c]{@{}c@{}}spy\\ Counter\end{tabular} &
  \multicolumn{1}{c|}{3 bits} &
  \begin{tabular}[c]{@{}c@{}}3*4096\\ =1.5 KB\end{tabular} &
  \multicolumn{1}{c|}{3 bits} &
  \begin{tabular}[c]{@{}c@{}}3*4096\\ =1.5 KB\end{tabular} \\ \hline
  
\begin{tabular}[c]{@{}c@{}}trojan\\ Counter\end{tabular} &
  \multicolumn{1}{c|}{3 bits} &
  \begin{tabular}[c]{@{}c@{}}3*4096\\ =1.5 KB\end{tabular} &
  \multicolumn{1}{c|}{3 bits} &
  \begin{tabular}[c]{@{}c@{}}3*4096\\ =1.5 KB\end{tabular} \\ \hline
  
\textbf{\begin{tabular}[c]{@{}c@{}}Total \\ Overhead\end{tabular}} &
  \multicolumn{1}{c|}{\textbf{11 bits}} &
  \textbf{5.5 KB} &
  \multicolumn{1}{c|}{\textbf{39 bits}} &
  \textbf{19.5 KB} \\ \hline
\end{tabular}
\end{table}

\section{Conclusion}
\label{sec:con}
This paper proposes an effective and efficient mitigation mechanism, TPPD, for cross-core cache timing channel attacks. TPPD implements way-wise partitioning on the cache sets used for covert channel attacks but only for suspicious process pairs. However, remaining benign processes have unrestricted access to these and other sets, reducing the performance impact on system performance. It successfully abolishes LLC based covert communication between trojan and spy. Experiments have shown that it does not have any significant impact on the performance of benign applications (Parsec benchmark). The total storage overhead required for implementing TPPD design is approximately $\approx 0.26$\% of LLC size. Compared to the existing partitioning based attack prevention mechanism NOMO, it caused $23$\% less LLC misses.

\bibliographystyle{IEEEtran}
\bibliography{Bibliography}

\begin{thebibliography}{10}
\providecommand{\url}[1]{#1}
\csname url@samestyle\endcsname
\providecommand{\newblock}{\relax}
\providecommand{\bibinfo}[2]{#2}
\providecommand{\BIBentrySTDinterwordspacing}{\spaceskip=0pt\relax}
\providecommand{\BIBentryALTinterwordstretchfactor}{4}
\providecommand{\BIBentryALTinterwordspacing}{\spaceskip=\fontdimen2\font plus
\BIBentryALTinterwordstretchfactor\fontdimen3\font minus
  \fontdimen4\font\relax}
\providecommand{\BIBforeignlanguage}[2]{{%
\expandafter\ifx\csname l@#1\endcsname\relax
\typeout{** WARNING: IEEEtran.bst: No hyphenation pattern has been}%
\typeout{** loaded for the language `#1'. Using the pattern for}%
\typeout{** the default language instead.}%
\else
\language=\csname l@#1\endcsname
\fi
#2}}
\providecommand{\BIBdecl}{\relax}
\BIBdecl

\bibitem{percival}
C.~Percival, ``{Cache Missing For Fun And Profit},'' in \emph{In Proc. of
  BSDCan}, 2005.

\bibitem{efficient}
E.~Tromer, D.~A. Osvik, and A.~Shamir, ``{Efficient Cache Attacks on AES, and
  Countermeasures},'' \emph{Journal of Cryptology}, vol.~23, no.~1, pp. 37--71,
  2010.

\bibitem{cca-attack}
D.~A. Osvik, A.~Shamir, and E.~Tromer, ``{Cache Attacks and Countermeasures:
  The Case of AES},'' in \emph{Cryptographers’ track at the RSA conference},
  2006, pp. 1--20.

\bibitem{cache-time}
D.~J. Bernstein, ``{Cache-Timing Attacks on AES},'' 2005.

\bibitem{lastlevelcachesidechannels}
F.~Liu, Y.~Yarom, Q.~Ge, G.~Heiser, and R.~B. Lee, ``{Last-Level Cache
  Side-Channel Attacks are Practical},'' in \emph{Symp. on Security and
  Privacy}, 2015, pp. 605--622.

\bibitem{survey-jas}
J.~Kaur and S.~Das, ``A survey on cache timing channel attacks for multicore
  processors,'' \emph{Journal of Hardware and Systems Security}, pp. 1--21,
  2021.

\bibitem{survey-prabhat}
Y.~Lyu and P.~Mishra, ``A survey of side-channel attacks on caches and
  countermeasures,'' \emph{Journal of Hardware and Systems Security}, vol.~2,
  no.~1, pp. 33--50, 2018.

\bibitem{aes}
V.~Rijmen and J.~Daemen, ``{Advanced Encryption Standard},'' \emph{Federal
  Information Processing Standards Publications, National Institute of
  Standards and Technology}, pp. 19--22, 2001.

\bibitem{rsa}
R.~L. Rivest, A.~Shamir, and L.~Adleman, ``{A Method for Obtaining Digital
  Signatures and Public-key Cryptosystems},'' \emph{Communications of the ACM},
  vol.~21, no.~2, pp. 120--126, 1978.

\bibitem{ecdsa}
D.~Johnson, A.~Menezes, and S.~Vanstone, ``The elliptic curve digital signature
  algorithm,'' \emph{International journal of information security}, vol.~1,
  no.~1, pp. 36--63, 2001.

\bibitem{tracedriven}
O.~Ac{\i}i{\c{c}}mez and {\c{C}}.~K. Ko{\c{c}}, ``{Trace-Driven Cache Attacks
  on AES (Short Paper)},'' in \emph{Information and Communications Security},
  2006, pp. 112--121.

\bibitem{cachegames}
D.~Gullasch, E.~Bangerter, and S.~Krenn, ``{Cache Games--Bringing Access-Based
  Cache Attacks on AES to Practice},'' in \emph{IEEE Symp. on Security and
  Privacy}, 2011, pp. 490--505.

\bibitem{FairSDP}
S.~{Sari}, O.~{Demir}, and G.~{Kucuk}, ``{FairSDP: Fair and Secure Dynamic
  Cache Partitioning},'' in \emph{4\textsuperscript{th} Intl. Conf. on Computer
  Science and Engineering}, 2019, pp. 469--474.

\bibitem{cotsknight}
F.~Yao, H.~Fang, M.~Doroslovacki, and G.~Venkataramani, ``{COTSknight:
  Practical Defense against Cache Timing Channel Attacks using Cache Monitoring
  and Partitioning Technologies},'' in \emph{Hardware Oriented Security and
  Trust}.

\bibitem{ucp}
M.~K. Qureshi and Y.~N. Patt, ``Utility-based cache partitioning: A
  low-overhead, high-performance, runtime mechanism to partition shared
  caches,'' in \emph{2006 39th Annual IEEE/ACM International Symposium on
  Microarchitecture (MICRO'06)}.\hskip 1em plus 0.5em minus 0.4em\relax IEEE,
  2006, pp. 423--432.

\bibitem{partition-halwe}
P.~D. Halwe, S.~Das, and H.~K. Kapoor, ``Towards a better cache utilization
  using controlled cache partitioning,'' in \emph{2013 IEEE 11th International
  Conference on Dependable, Autonomic and Secure Computing}.\hskip 1em plus
  0.5em minus 0.4em\relax IEEE, 2013, pp. 179--186.

\bibitem{partition-new}
C.~Yang, L.~Liu, K.~Luo, S.~Yin, and S.~Wei, ``Ciacp: A correlation-and
  iteration-aware cache partitioning mechanism to improve performance of
  multiple coarse-grained reconfigurable arrays,'' \emph{IEEE Transactions on
  Parallel and Distributed Systems}, vol.~28, no.~1, pp. 29--43, 2016.

\bibitem{anuraag}
A.~Agarwal, J.~Kaur, and S.~Das, ``Exploiting secrets by leveraging dynamic
  cache partitioning of last level cache,'' in \emph{Design, Automation Test in
  Europe Conference Exhibition (DATE)}, 2021.

\bibitem{prodact}
H.~Fang, S.~S. Dayapule, F.~Yao, M.~Doroslova{\v{c}}ki, and G.~Venkataramani,
  ``{Prodact: Prefetch-obfuscator to Defend against Cache Timing Channels},''
  \emph{Intl. Journal of Parallel Programming}, vol.~47, no.~4, pp. 571--594,
  2019.

\bibitem{parsec}
C.~Bienia, \emph{Benchmarking modern multiprocessors}.\hskip 1em plus 0.5em
  minus 0.4em\relax Princeton University, 2011.

\bibitem{gem5}
N.~Binkert, B.~Beckmann, G.~Black, S.~K. Reinhardt, A.~Saidi, A.~Basu,
  J.~Hestness, D.~R. Hower, T.~Krishna, S.~Sardashti \emph{et~al.}, ``The gem5
  simulator,'' \emph{ACM SIGARCH computer architecture news}, vol.~39, no.~2,
  pp. 1--7, 2011.

\bibitem{basicevictionset}
G.~{Irazoqui}, T.~{Eisenbarth}, and B.~{Sunar}, ``{S\$A: A Shared Cache Attack
  That Works across Cores and Defies VM Sandboxing -- and Its Application to
  AES},'' in \emph{IEEE Symp. on Security and Privacy}, 2015, pp. 591--604.

\bibitem{newcachedesigns}
Z.~Wang and R.~B. Lee, ``{New Cache Designs for Thwarting Software Cache-based
  Side Channel Attacks},'' \emph{SIGARCH Comput. Archit. News}, vol.~35, no.~2,
  pp. 494--505, 2007.

\bibitem{nomo}
L.~Domnitser, A.~Jaleel, J.~Loew, N.~Abu-Ghazaleh, and D.~Ponomarev,
  ``{Non-Monopolizable Caches: Low-Complexity Mitigation of Cache Side Channel
  Attacks},'' \emph{Transactions on Architecture and Code Optimization},
  vol.~8, no.~4, pp. 1--21, 2012.

\bibitem{secdcp}
Y.~Wang, A.~Ferraiuolo, D.~Zhang, A.~C. Myers, and G.~E. Suh, ``Secdcp: Secure
  dynamic cache partitioning for efficient timing channel protection,'' in
  \emph{2016 53nd ACM/EDAC/IEEE Design Automation Conference (DAC)}, 2016, pp.
  1--6.

\bibitem{ceaser}
M.~K. Qureshi, ``Ceaser: Mitigating conflict-based cache attacks via
  encrypted-address and remapping,'' in \emph{2018 51st Annual IEEE/ACM
  International Symposium on Microarchitecture (MICRO)}, 2018, pp. 775--787.

\bibitem{ceaser-s}
------, ``New attacks and defense for encrypted-address cache,'' in
  \emph{Proceedings of the 46th International Symposium on Computer
  Architecture}, ser. ISCA '19.\hskip 1em plus 0.5em minus 0.4em\relax New
  York, NY, USA: Association for Computing Machinery, 2019, p. 360–371.

\bibitem{damru}
P.~Kumar, C.~S. Yashavant, and B.~Panda, ``Damaru: A denial-of-service attack
  on randomized last-level caches,'' \emph{IEEE Computer Architecture Letters},
  vol.~20, no.~2, pp. 138--141, 2021.

\bibitem{hawkeye}
A.~Jain and C.~Lin, ``Back to the future: Leveraging belady's algorithm for
  improved cache replacement,'' in \emph{Proceedings of the 43rd International
  Symposium on Computer Architecture}, ser. ISCA '16.\hskip 1em plus 0.5em
  minus 0.4em\relax IEEE Press, 2016, p. 78–89.

\bibitem{negative-corelation}
H.~Fang, F.~Yao, M.~Doroslovački, and G.~Venkataramani, ``Negative
  correlation, non-linear filtering, and discovering of repetitiveness for
  cache timing channel detection,'' in \emph{ICASSP 2019 - 2019 IEEE
  International Conference on Acoustics, Speech and Signal Processing
  (ICASSP)}, 2019, pp. 2522--2526.

\bibitem{rrip}
A.~Jaleel, K.~B. Theobald, S.~C. Steely~Jr, and J.~Emer, ``High performance
  cache replacement using re-reference interval prediction (rrip),'' \emph{ACM
  SIGARCH Computer Architecture News}, vol.~38, no.~3, pp. 60--71, 2010.

\bibitem{prathamesh}
K.~K. Dutta, P.~N. Tanksale, and S.~Das, ``A fairness conscious cache
  replacement policy for last level cache,'' in \emph{2021 Design, Automation
  \& Test in Europe Conference \& Exhibition (DATE)}.\hskip 1em plus 0.5em
  minus 0.4em\relax IEEE, 2021, pp. 695--700.

\bibitem{warrier2013application}
T.~S. Warrier, B.~Anupama, and M.~Mutyam, ``An application-aware cache
  replacement policy for last-level caches,'' in \emph{International Conference
  on Architecture of Computing Systems}.\hskip 1em plus 0.5em minus 0.4em\relax
  Springer, 2013, pp. 207--219.

\bibitem{parsec2009}
C.~Bienia and K.~Li, ``Parsec 2.0: A new benchmark suite for
  chip-multiprocessors,'' in \emph{Proceedings of the 5th Annual Workshop on
  Modeling, Benchmarking and Simulation}, vol. 2011, 2009, p.~37.

\bibitem{secverilog}
D.~Zhang, Y.~Wang, G.~E. Suh, and A.~C. Myers, ``A hardware design language for
  timing-sensitive information-flow security,'' \emph{Acm Sigplan Notices},
  vol.~50, no.~4, pp. 503--516, 2015.

\end{thebibliography}

\begin{IEEEbiography}[{\includegraphics[width=1in,height=1.25in,clip,keepaspectratio]{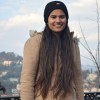}}]%
{Jaspinder Kaur}
A Ph.D scholar in the CSE department of Indian Institute of Technology Ropar. She has completed M.Tech (CSE) from the Department of Computer Engineering of Punjabi University Patiala, India. Her research interests include Computer Architecture, Cache Prefetching and Cache Security against timing channel attacks.
\end{IEEEbiography}
\begin{IEEEbiography}[{\includegraphics[width=1in,height=1.25in,clip,keepaspectratio]{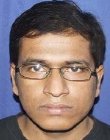}}]%
{Shirshendu Das}
Received a Ph.D. degree (CSE) from the Indian Institute of Technology Guwahati, India, in 2016. Previously he did M.Tech (CSE) from the Indian Institute of Technology Guwahati, India. Presently he is an Assistant Professor in the Department of CSE, Indian Institute of Technology Ropar, Punjab, India. His area of research includes Computer Architecture, Network-on-Chip, and Hardware Security.
\end{IEEEbiography}
\end{document}